\documentclass[a4paper,11pt]{article}
\usepackage{pos}
\usepackage{graphicx}
\usepackage{tabularx}
\usepackage{slashed}
\usepackage{mathtools}

\newcommand{\Q}{\mathbb{Q}}

\renewcommand\phi{\varphi}

\definecolor{darkgray}{rgb}{0.3,0.3,0.3}
\definecolor{LightGray}{gray}{0.9}

\def\bea{\begin{eqnarray}}
\def\eea{\end{eqnarray}}

\def\be{\begin{equation}}
\def\ee{\end{equation}}

\title{IBP reduction via Gr\"obner bases \\ in a rational double-shift algebra}

\ShortTitle{IBP reduction via Gr\"obner bases in a rational double-shift algebra}

\author[a]{Mohamed Barakat}
\author[b]{Robin Br\"user}
\author*[b]{Tobias Huber}
\author[b]{Jan Piclum}

\affiliation[a]{Department Mathematik, Naturwissenschaftlich-Technische Fakult\"at, Universit\"at Siegen, \\
Walter-Flex-Str.~3, 57068 Siegen,Germany}

\affiliation[b]{Theoretische Physik 1, Center for Particle Physics Siegen (CPPS), Universit\"at Siegen, \\
Walter-Flex-Str.~3, 57068 Siegen, Germany}

\emailAdd{mohamed.barakat@uni-siegen.de}
\emailAdd{robin.brueser@uni-siegen.de}
\emailAdd{huber@physik.uni-siegen.de}
\emailAdd{piclum@physik.uni-siegen.de}

\abstract{We report on an approach to integration-by-parts reduction based on Gr\"obner bases. We establish the underlying noncommutative rational double-shift algebra wherein the integration-by-parts relations form a left ideal. We describe in detail the one-loop massless box as an example where we achieved the full reduction to master integrals by means of the Gr\"obner basis approach, and report on the performance of the implementation. We also identify potential bottlenecks in more complicated examples and elaborate on interesting further directions.
}

\FullConference{%
  Loops and Legs in Quantum Field Theory - LL2022,\\
  25-30 April, 2022\\
  Ettal, Germany
}

\begin{document}

\renewcommand{\hookAfterAbstract}{%
\par\bigskip
\textsc{SI-HEP-2022-16, P3H-22-078, ArXiv ePrint}:
\href{https://arxiv.org/abs/2207.09275}{2207.09275}
}
\maketitle

\section{Introduction}
\label{sec:intro}

Reduction of dimensionally regularized Feynman integrals based on integration-by-parts (IBP) relations~\cite{Tkachov:1981wb,Chetyrkin:1981qh} is an indispensable tool for carrying out higher-order calculations in perturbative quantum field theory. Many sophisticated public and private codes to perform this task exist in various programming languages, for instance ${\mathtt{AIR}}$~\cite{Anastasiou:2004vj}, ${\mathtt{FIRE}}$~\cite{Smirnov:2008iw,Smirnov:2014hma,Smirnov:2019qkx}, 
${\mathtt{Reduze}}$~\cite{Studerus:2009ye,vonManteuffel:2012np}, ${\mathtt{LiteRed}}$~\cite{Lee:2012cn}, and ${\mathtt{Kira}}$~\cite{Maierhofer:2017gsa,Klappert:2020nbg}.

The reduction procedures that have been implemented in these programs are mostly\footnote{${\mathtt{LiteRed}}$ instead uses a heuristic which provides symbolic rules valid for the reduction of any integral of the family.} based on Laporta's algorithm~\cite{Laporta:2000dsw} which solves the IBP equations for numerical values of the propagator powers of the integral using ``bottom-up'' Gaussian elimination. In recent years, several refinements of this algorithm have been developed to speed up the calculation, which are mostly based on parallelization and ideas from finite fields and rational reconstruction~\cite{vonManteuffel:2014ixa,Peraro:2016wsq,Smirnov:2019qkx,Peraro:2019svx,Klappert:2019emp,Klappert:2020aqs}.

The Laporta algorithm has served the community in countless multi-loop calculations over the past two decades. It has, however, also a couple of drawbacks. For instance, in many cases redundant integrals have to be computed during the reduction procedure in order to get access to those integrals that are required by the actual calculation of physical quantities. Consequently, storing the results of typically ${\cal O}(10^{\sim 4-6})$ integrals demands for large storage capacities. Moreover, plugging in integer values for the propagator powers generates a huge system of equations, whose solution via Gaussian elimination generates a considerable expression swell at intermediate stages, at least as long as none of the aforementioned refinements are applied.

More recently, new ideas towards a more direct reduction procedure have been developed. They are mostly based on syzygy equations~\cite{Gluza:2010ws,Schabinger:2011dz,Lee:2014tja,Bohm:2017qme,Kosower:2018obg}, algebraic geometry~\cite{Larsen:2015ped,Bohm:2018bdy,Bendle:2019csk}, and intersection numbers~\cite{Mastrolia:2018uzb,Frellesvig:2019uqt,Frellesvig:2019kgj,Abreu:2019wzk,Weinzierl:2020xyy,Caron-Huot:2021iev,Chen:2022lzr,Chestnov:2022alh}.
In these proceedings we report on work in progress~\cite{BBFHP}, where we choose an approach to IBP reduction that is based on Gr\"obner bases and hence leaves the propagator powers parametric. While IBP reductions by means of Gr\"obner bases have been attempted in the past~\cite{Tarasov:1998nx,Tarasov:2004ks,Gerdt:2004kt,Gerdt:2005qf,Smirnov:2005ky,Smirnov:2006tz,Smirnov:2006wh,Lee:2008tj}, we formulate for the first time the appropriate noncommutative rational double-shift algebra wherein the IBP relations generate a left ideal. For selected examples of which we describe one representative below, we were able to compute the Gr\"obner basis for the left ideal of IBP relations in the noncommutative rational double-shift algebra and achieved the full reduction with the Gr\"obner basis technique.

This article is organized as follows.
In the next section we recap the basics about Gr\"obner bases and related terms from algebraic geometry.
In section~\ref{sec:Y} we establish the noncommutative rational double-shift algebra wherein the IBP relations form a left ideal.
Section~\ref{sec:1LB} contains the one-loop massless box as an explicit example where we achieved a full reduction with the Gr\"obner basis approach. We conclude in section~\ref{sec:conc}.

\section{Basics about Gr\"obner bases}
\label{sec:GBbas}

We first give the definitions of a few key quantities necessary for our calculation and its description in the subsequent sections.

Let $R$ be a ring. A {\textit{left ideal}} $I \subseteq R$ is an additive subgroup of $R$ fulfilling
\begin{equation}\label{eq:ideal}
r \in R \wedge a \in I  \; \Longrightarrow \; r a \in I.
\end{equation}
As a simple example, the set of even integers forms a (left) ideal in the ring $\mathbb{Z}$ of integers.

A {\textit{monomial order}} on the polynomial algebra $R = \mathbb{K}[x] = \mathbb{K}[x_1, \ldots, x_n]$ over a field $\mathbb{K}$ is a total order $>$ such that
\begin{equation}\label{eq:monord}
x^\alpha > x^\beta  \; \Longrightarrow \; x^\gamma x^\alpha > x^\gamma x^\beta \quad \forall \, \alpha,\beta,\gamma \in \mathbb{N}^n \, ,
\end{equation}
where $\alpha$, $\beta$, and $\gamma$ are multi-indices. The most prominent (global) monomial orders are the {\textit{lexicographic order}} for which
\begin{equation}\label{eq:lex}
x^\alpha >_{\mbox{{\scriptsize{lex}}}} x^\beta \quad \Longleftrightarrow \quad  \mbox{first nonzero entry of } \alpha - \beta >0
\end{equation}
has to hold, and the {\textit{degree reverse lexicographic order}} with condition
\begin{equation}\label{eq:degrevlex}
x^\alpha >_{\mbox{{\scriptsize{drlex}}}} x^\beta  \Longleftrightarrow  (\mbox{\footnotesize{deg} } x^\alpha > \mbox{\footnotesize{deg} } x^\beta) \mbox{ or } (\mbox{\footnotesize{deg} } x^\alpha = \mbox{\footnotesize{deg} } x^\beta \mbox{ and last nonzero entry of } \alpha - \beta <0).
\end{equation}

For $f\in R$, the {\textit{leading term}} ${\boldsymbol{L}}_>(f)$ with respect to a given monomial order $>$ is the largest term in $f$ with respect to $>$.
A finite subset $G = \{f_1,\ldots,f_r\} \subset I$ is a {\textit{Gr\"obner basis for the (left) ideal $I$}} if
\begin{equation}\label{eq:defGB}
{\boldsymbol{L}}_>(I) = {\boldsymbol{L}}_>(G) \, ,
\end{equation}
i.e.\ the leading (left) ideal of $I$ is generated by the leading terms of the elements of $G$. Hence $G$ generates $I$. One way of computing Gr\"obner bases in polynomial algebras is via Buchberger's algorithm.
In this work we use a generalization of Buchberger's algorithm to the context of Ore algebras as developed in~\cite{Chyzak-1998-GBS}.
This class includes the aforementioned polynomial algebras, but also a wide class of noncommutative algebras, including the rational double-shift algebra which is central to this work.

The {\textit{remainder}} $h$ of $g = \sum_{i=1}^r \, g_i \, f_i + h$ is uniquely determined by $g$, $I$, and $>$.
Moreover, we will call $\operatorname{NF}_{I,>}(g) = \operatorname{NF}_G(g) = h$ the {\textit{normal form}} of $g$ mod $I$ with respect to $>$.

\section{Noncommutative rational double-shift algebra}
\label{sec:Y}

We start from a generic $L$-loop integral
\begin{equation}
  J(a_1, \ldots, a_n)
  =
  \int
  \operatorname{d}^D \ell_1 \cdots \operatorname{d}^D \ell_L \,
  \frac{1}{P_1^{a_1} \cdots P_n^{a_n}} \mbox{,}
\end{equation}
where $D$ is the number of space-time dimensions in dimensional regularization. Each of the propagators $P_i$, $i=1,\ldots,n$, is usually of the form $P_i = m_i^2-p_i^2$ with mass $m_i$ and $p_i$ a linear combination of the $L$ loop momenta $\ell_1, \ldots, \ell_L$ and $E$ external momenta $k_1, \ldots, k_E$.
The integral therefore depends on the propagator powers (indices) $a_i$, the number of space-time dimensions $D$, the masses $m_i^2$ and kinematic invariants built out of the the external momenta which we collectively label $s_{ij}$. In the following we will suppress all dependence of $J$ but that on the indices $a_i$.

The $L(L+E)$ standard IBP relations that are derived from
\begin{align}
  \int\operatorname{d}^D \ell_1 \cdots \operatorname{d}^D \ell_L
  \; \frac{\partial}{\partial \ell_j^\mu} \left( 
  \frac{v_k^\mu}{P_1^{a_1} \cdots P_n^{a_n}} \right) & = 0 \, \mbox{,} \label{eq:IBP}
\end{align}
with $v_k^\mu$ any loop or external momentum, can be expressed in terms of shift operators $D_i$, $D_i^-$ and multiplication operators $a_i$, $i=1,\ldots,n$, with the following partial right action on the space of loop integrals $J(z_1,\ldots,z_n)$:
\begin{align}
  J(\ldots,z_i,\ldots) \bullet D_i &= J(\ldots,z_i-1,\ldots), &\quad  \underbrace{J(\ldots,z_i,\ldots)}_{\text{not scaleless}} \bullet D_i^- &= J(\ldots,z_i+1,\ldots),
  \nonumber\\ 
  J(\ldots,z_i,\ldots) \bullet a_i &= z_i J(\ldots,z_i,\ldots), &\quad J(\ldots,\underbrace{z_i}_{\neq 0},\ldots) \bullet a_i^{-1} &= \frac{1}{z_i} J(\ldots,z_i,\ldots) \mbox{.}
\end{align}
Our computations take place in the noncommutative rational double-shift algebra
\begin{equation}
 Y := \mathbb{Q}(D,s_{ij},m_i^2)(a_1,\ldots,a_n)\langle D_j,D_j^- \mid j=1,\ldots,n\rangle 
\end{equation}
in the indeterminates $a_1, \ldots, a_n, D_1, \ldots, D_n, D_1^-, \ldots, D_n^-$ which satisfy the relations\footnote{no summation over repeated indices}
\begin{align}
  [a_i, D_j] = \delta_{ij} D_i \, \mbox{,}\qquad [a_i, D_j^-] = -\delta_{ij}  D_i^- \, \mbox{,} \qquad D_i D_i^- = 1 = D_i^- D_i \, \mbox{,}
  \nonumber\\[0.2em]
  [a_i, a_j] = [D_i, D_j] = [D_i^-, D_j^-] = [D_i, D_j^-] = 0 \, \mbox{.}
\end{align}
The standard IBP relations generate a left ideal in the noncommutative rational double-shift algebra
\begin{align*}
  I_\mathrm{IBP} &\coloneqq \langle r_i \mid i = 1, \ldots, L(L+E) \rangle_Y \lhd Y \mbox{.}
\end{align*}
Our goal will be to compute a Gr\"obner basis for the left ideal $I_\mathrm{IBP}$ in $Y$.

We close this section by defining a {\textit{standard monomial}} with respect to the Gr\"obner basis $G$ of $I_\mathrm{IBP}$, which is a monomial $m$ in the indeterminates $D_i, D_j^-$ such that $\operatorname{NF}_G(m) = m$. The set of standard monomials forms a basis for the finite-dimensional vector space $Y / I_\mathrm{IBP}$ over the field $\mathbb{Q}(D,s_{ij},m_i^2)$ of coefficients, and corresponds to a set of master integrals with respect to some fixed initial integral, usually taken to be the corner integral of the topology under consideration.

For the technical implementation we developed the {\textsf{GAP}} package $\mathtt{LoopIntegrals}$~\cite{LoopIntegrals}, which relies  on Chyzak's {\textsf{Maple}} package $\mathtt{Ore\_algebra}$~\cite{Chyzak-1998-GBS} to perform Gr\"obner basis computations in the noncommutative double-shift algebra with rational coefficients.
The interface between {\textsf{GAP}}~\cite{GAP4111} and  $\mathtt{Ore\_algebra}$ is provided by the $\mathtt{homalg}$-project packages~\cite{homalg-project}.

\section{One-loop massless box}
\label{sec:1LB}

\begin{figure}
  \begin{center}
    \includegraphics[width=0.3\textwidth]{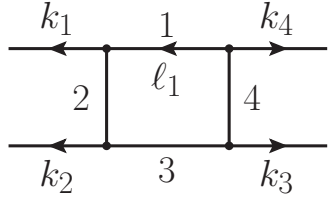}
  \end{center}
  \caption{\label{fig:olb}The Feynman graph of the one-loop box integral.}
\end{figure}

An example of a successful application of the Gr\"obner basis approach to IBP reduction is the one-loop massless box depicted in figure~\ref{fig:olb}. It is defined by the loop momentum $\ell_1$ and the external momenta $k_1, k_2, k_3, k_4$, of which we take 
$k_1, k_2, k_4$ to be the linearly independent ones. The external lines are on-shell and massless, i.e.\ $k_i^2=0$ for $i=1,2,3,4$, which results in the independent external kinematic invariants $s_{12} = 2 k_1 \cdot k_2$ and $s_{14} = 2 k_1 \cdot k_4$. Internal lines are also massless. The $n=4$ propagators are
\begin{align}
  P_1 &= -\ell_1^2 \, \mbox{,} &
  P_2 &= -(\ell_1-k_1)^2 \, \mbox{,} &
  P_3 &= -(\ell_1 - k_1 - k_2)^2 \, \mbox{,} &
  P_4 &= -(\ell_1 + k_4)^2 \, \mbox{,}
\end{align}
from which we derive the four standard IBP relations
{\small
\begin{align}
  r_1
  &= -a_2 D_1 D_2^- -a_3 D_1 D_3^- -a_4 D_1 D_4^- -s_{12} a_3 D_3^- + (D-2a_1-a_2-a_3-a_4), \nonumber\\
  r_2
  &= a_1 D_1^- D_2 - a_2 D_1 D_2^- -a_3 D_1 D_3^- + a_3 D_2 D_3^- - a_4 D_1 D_4^- + a_4 D_2 D_4^- -s_{12} a_3 D_3^- + s_{14} a_4 D_4^- -a_1+a_2, \nonumber\\
  r_3
  &= - a_1 D_1^- D_2 + a_1 D_1^- D_3 + a_2 D_2^- D_3 - a_3 D_2 D_3^- - a_4 D_2 D_4^- + a_4 D_3 D_4^- + s_{12} a_1 D_1^- - s_{14} a_4 D_4^- -a_2 + a_3, \nonumber\\
  r_4
  &= a_2 D_1 D_2^- + a_3 D_1 D_3^- - a_1 D_1^- D_4 - a_2 D_2^- D_4 - a_3 D_3^- D_4 + a_4 D_1 D_4^- - s_{14} a_2 D_2^- + s_{12} a_3 D_3^- +a_1-a_4 \mbox{.}
  \label{eq:standardIBP}
\end{align}
}
 
By means of the techniques described in the previous sections, we compute the reduced Gr\"obner basis $G$ in the noncommutative rational double-shift algebra
\begin{equation}
    Y = \mathbb{K}(a_1,a_2,a_3,a_4) \langle D_i, D_i^- \mid i = 1, \ldots, 4 \rangle \, \mbox{.}
\end{equation}
over the field $\mathbb{K} = \Q(D,s_{12},s_{14})$ of coefficients.
 It has the 9 elements
 \allowdisplaybreaks
    \begin{align}
    \allowdisplaybreaks
    G&
    = \bigg\{ D_4-D_2+\frac{(a_2-a_4) s_{14}}{D-a_{1234}}, D_3-D_1+\frac{(a_1-a_3) s_{12}}{D-a_{1234}}, \nonumber \\[0.4em]
 & 4 (a_2-1) (D-a_{1234}) D_3-2 (D-2a_{134}) (D-a_{1234}) D_4 +(D-2a_{14}-2) (D-2a_{234}) s_{12}  \nonumber\\[0.2em]
 & -2 (D-2a_{134}) (a_2-a_4) s_{14} -\frac{(D-2a_{14}-2) (D-2a_{34}-2) a_4 s_{12} s_{14}}{D-a_{1234}-1} \, D_4^{-}, \nonumber \\[0.4em]
 & -2 (D-2a_{234}) (D-a_{1234}) D_3+4 (a_1-1) (D-a_{1234}) D_4-2 (a_1-a_3) (D-2a_{234}) s_{12} \nonumber\\[0.2em]
 & +(D-2a_{23}-2) (D-2a_{134}) s_{14} -\frac{(D-2a_{23}-2) a_3 (D-2a_{34}-2) s_{12} s_{14}}{D-a_{1234}-1} \, D_3^{-}, \nonumber \\[0.4em]
 & 4 (D-a_{1234}) (a_4-1) D_3-2 (D-2a_{123}) (D-a_{1234}) D_4\nonumber\\[0.2em]
 & +(D-2a_{12}-2) (D-2a_{234}) s_{12} -\frac{(D-2a_{12}-2) a_2 (D-2a_{23}-2) s_{12} s_{14}}{D-a_{1234}-1} \, D_2^{-}, \nonumber \\[0.4em]
 & -2 (D-2a_{124}) (D-a_{1234}) D_3+4 (a_3-1) (D-a_{1234}) D_4\nonumber\\[0.2em]
 &+(D-2a_{12}-2) (D-2a_{134}) s_{14} -\frac{a_1 (D-2a_{12}-2) (D-2a_{14}-2) s_{12} s_{14}}{D-a_{1234}-1} \, D_1^{-}, \nonumber \\[0.4em]
 & -2 (D-2a_{1234}+4) (D-a_{1234}+1) D_4^2 +(D-2a_{124}+2) (D-2a_{234}+2) s_{12} D_4\nonumber\\[0.2em]
 &-2 (D-2a_{1234}+4) (a_2-a_4+1) s_{14} D_4+4 (a_2-1) (a_4-1) s_{14} D_3\nonumber\\[0.2em]
 & -\frac{(D-2a_{124}+2) (D-2a_{34}) (a_4-1) s_{12} s_{14}}{D-a_{1234}}, \nonumber \\[0.4em]
 & -(D-2a_{1234}+4) (D-a_{1234}+1) D_3 D_4 +(a_3-1) (D-2a_{234}+2) s_{12} D_4 \nonumber\\[0.2em]
 &+(D-2a_{134}+2) (a_4-1) s_{14} D_3 -\frac{(a_3-1) (D-2a_{34}) (a_4-1) s_{12} s_{14}}{D-a_{1234}}, \nonumber \\[0.4em]
 & -2 (D-2a_{1234}+4) (D-a_{1234}+1) D_3^2 +(D-2a_{123}+2) (D-2a_{134}+2) s_{14} D_3\nonumber\\[0.2em]
 & - 2 (a_1-a_3+1) (D-2a_{1234}+4) s_{12} D_3+4 (a_1-1) (a_3-1) s_{12} D_4 \nonumber\\[0.2em]
 &-\frac{(D-2a_{123}+2) (a_3-1) (D-2a_{34}) s_{12} s_{14}}{D-a_{1234}}\bigg\} \mbox{,} \label{eq:GB}
  \end{align}
  with the abbreviations $a_{i_1 \ldots i_k} \coloneqq \sum_{j=1}^k a_{i_j}$.
  The Gröbner basis $G$ is rational in $D$, $a_i$, $s_{ij}$, and polynomial in $D_i$ and $D_i^-$, as expected.

We can now compute the normal forms of the indeterminates $D_i$, which reveal the $V_4$-symmetry of the problem,
  \begin{align}
    \operatorname{NF}_G(D_1) =
    &
    D_3+\frac{(a_1-a_3)s_{12}}{D-a_{1234}} \, ,
    &\operatorname{NF}_G(D_3) = & D_3 \, ,
    \nonumber\\
    \operatorname{NF}_G(D_2) =
    &
    D_4+\frac{(a_2-a_4)s_{14}}{D-a_{1234}} \, ,
    &\operatorname{NF}_G(D_4) = & D_4 \, . \label{eq:NFD}
  \end{align}
The set of standard monomials with respect to $G$ is therefore $\{ 1, D_3, D_4 \}$, which correspond to the three master integrals
  \begin{align} \label{eq:masters_oneloopbox}
    \{ J(1,1,1,1), J(1,1,0,1), J(1,1,1,0) \} \mbox{,}
  \end{align}
i.e.\ the Gr\"obner basis reduction yields the box and two triangles as basis of master integrals.
Computing further the normal forms of the monomials $D_1 D_2$, $D_1 D_4$, $D_2 D_3$, $D_3 D_4$ with respect to the Gr\"obner basis $G$, one can easily verify that they are scaleless with respect to $J(1,1,1,1)$.
Based on this and other examples we conjecture that the Gr\"obner basis reduction recognizes the scaleless integrals of a given topology.

We proceed by computing the normal form of the operators $a_i D_i^-$ with respect to the Gr\"obner basis $G$ of the left ideal
$ I_\mathrm{IBP} = \langle r_i \mid i = 1, \ldots, 4 \rangle_Y \lhd Y$
generated by the standard IBP relations in eq.~(\ref{eq:standardIBP}),
   \allowdisplaybreaks
  {\small 
   \allowdisplaybreaks
  \begin{align}
  \allowdisplaybreaks
    \operatorname{NF}_G(a_1 D_1^-) =
    &
    -\frac{2 \left(D-2a_{124}\right) \! \left(D-a_{1234}\right) \! \left(D-a_{1234}-1\right)}{\left(D-2a_{12}-2\right) \left(D-2a_{14}-2\right) s_{12} s_{14}} D_3 + \frac{4 \left(a_3-1\right) \! \left(D-a_{1234}\right) \!\left(D-a_{1234}-1\right)}{\left(D-2a_{12}-2\right) \left(D-2a_{14}-2\right) s_{12} s_{14}} D_4 \nonumber\\
   & + \frac{\left(D-2a_{134}\right) \left(D-a_{1234}-1\right)}{\left(D-2a_{14}-2\right) s_{12}},
    \nonumber\\
    \operatorname{NF}_G(a_2 D_2^-) =
    &
    \frac{4 \left(a_4-1\right) \left(D-a_{1234}\right) \left(D-a_{1234}-1\right)}{\left(D-2a_{12}-2\right)
   \left(D-2a_{23}-2\right) s_{12} s_{14}} D_3 -\frac{2 \left(D-2a_{123}\right) \left(D-a_{1234}\right) \left(D-a_{1234}-1\right)}{\left(D-2a_{12}-2\right) \left(D-2a_{23}-2\right) s_{12} s_{14}} D_4 \nonumber\\
   &+ \frac{\left(D-2a_{234}\right) \left(D-a_{1234}-1\right)}{\left(D-2a_{23}-2\right) s_{14}},
    \nonumber\\
    \operatorname{NF}_G(a_3 D_3^-) =
    &
    -\frac{2 \left(D-2a_{234}\right) \! \left(D-a_{1234}\right) \! \left(D-a_{1234}-1\right)}{\left(D-2 a_{23}-2\right) \left(D-2a_{34}-2\right) s_{12} s_{14}} D_3
    +\frac{4 \left(a_1-1\right) \! \left(D-a_{1234}\right) \! \left(D-a_{1234}-1\right)}{\left(D-2a_{23}-2\right) \left(D-2a_{34}-2\right) s_{12} s_{14}} D_4
    \nonumber\\
    &+\frac{\left(D-2a_{134}\right) \left(D-a_{1234}-1\right)}{\left(D-2a_{34}-2\right) s_{12}}-\frac{2 \left(a_1-a_3\right)
   \left(D-2a_{234}\right) \left(D-a_{1234}-1\right)}{\left(D-2a_{23}-2\right) \left(D-2a_{34}-2\right)
   s_{14}},
    \nonumber\\
    \operatorname{NF}_G(a_4 D_4^-) =
    &
    \frac{4 \left(a_2-1\right) \left(D-a_{1234}\right) \left(D-a_{1234}-1\right)}{\left(D-2a_{14}-2\right) \left(D-2a_{34}-2\right) s_{12} s_{14}} D_3
   -\frac{2 \left(D-2a_{134}\right) \left(D-a_{1234}\right) \left(D-a_{1234}-1\right)}{\left(D-2a_{14}-2\right) \left(D-2a_{34}-2\right) s_{12} s_{14}} D_4 \nonumber\\
    &
    +\frac{\left(D-2a_{234}\right) \left(D-a_{1234}-1\right)}{\left(D-2a_{34}-2\right) s_{14}}-\frac{2 \left(a_2-a_4\right) \left(D-2a_{134}\right) \left(D-a_{1234}-1\right)}{\left(D-2a_{14}-2\right) \left(D-2a_{34}-2\right) s_{12}} \mbox{.} \label{eq:NFone-loop-box}
  \end{align}
  }%
All $\operatorname{NF}_G(D_i)$ and $\operatorname{NF}_G(a_i D_i^-)$ are $\mathbb{K}$-linear combinations of the standard monomials, which leads us to conjecture that the Gr\"obner basis reduction also recognizes the symmetries of a problem. Moreover, we observe that in both equations~(\ref{eq:NFD}) and~(\ref{eq:NFone-loop-box}) no nonconstant polynomials in $\mathbb{Q}[a_1,\ldots,a_4]$ appear in the denominator, which means that these denominator factors in $\mathbb{K}[a_1,\ldots,a_4]$ never vanish within dimensional regularization.

We conclude this section by some information on runtimes for various parts of the calculation. The Gr\"obner basis in eq.~(\ref{eq:GB}) was computed in less than 5 seconds on a modern laptop. We also implemented the computation of normal forms modulo $G$ in a {\textsf{FORM}}~\cite{Ruijl:2017dtg} code which we will provide electronically with~\cite{BBFHP}. The {\textsf{FORM}} program is able to do fast reductions, even for rather large values of the indices. For instance, it expresses $J(10, 10, 10, 10)$ in terms of master integrals in less than 10 seconds on a desktop computer.
However, we also mention that for problems that look at first glance only slightly more complicated than the one-loop massless box, we observe an extraordinary swell in runtime and memory consumption when attempting to compute a Gr\"obner basis. We will give more details on this circumstance in the next section.

\section{Conclusion and outlook}
\label{sec:conc}

We reported on recent progress in the Gr\"obner basis approach to IBP reduction.
A key step towards a successful reduction of nontrivial Feynman integrals was to recognize that for our setup the noncommutative rational double-shift algebra is the proper algebra wherein the IBP relations generate a left ideal.
The computations are organized by means of the {\textsf{GAP}} package $\mathtt{LoopIntegrals}$, which relies on the noncommutative Gr\"obner basis algorithms provided by Chyzak's {\textsf{Maple}} package $\mathtt{Ore\_algebra}$.

We elaborated in detail on the one-loop massless box, for which we achieved a full reduction to master integrals within very short runtimes. This example also shows a number of appealing features of the Gr\"obner basis approach to IBP reduction. First, with the Gr\"obner basis at hand, the entire information required for reduction is available for any values of the propagator powers, which entails that no new bottom-up reduction is required if one seeks for the reduction of new or additional integrals of the same family. A second important feature that we observed in the example of the one-loop massless box is the recognition of symmetries and scaleless sectors of an integral family, which we conjecture to happen also for other, more complicated topologies. 

\begin{figure}
  \begin{center}
    \includegraphics[width=0.4\textwidth]{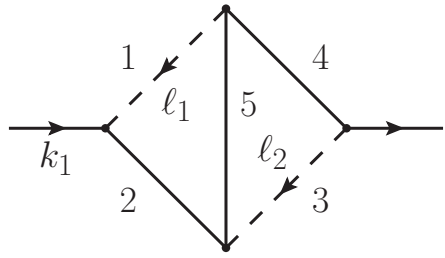}
  \end{center}
  \caption{\label{fig:on-shell-kite}The Feynman graph for the kite integral. Dashed lines denote massless propagators and solid lines
    denote massive propagators of mass $m$. The on-shell condition implies $k_1^2=m^2$.}
\end{figure}

However, as so often, there is no free lunch, and hence, as the complexity of the problem increases, the Gr\"obner basis technique also reveals bottlenecks which potentially eat up parts of the virtues identified above. Let's consider, for instance, the two-loop on-shell kite integral whose diagram is shown in figure~\ref{fig:on-shell-kite}. Compared to the one-loop massless box it has an extra loop but only a single scale. One might therefore expect the complexity of the kite to be moderately above that of the box. However, the computation of the Gr\"obner basis results in an extraordinary expression swell which as of now prevented us from finishing the computation. Still, we were able to compute the normal forms $\operatorname{NF}(a_i D_i^-)$, $i=1,\ldots,5$ using a linear algebra ansatz~\cite{BBFHP}, which allows, e.g., for the reduction of the top-level sector.

To conclude, the Gr\"obner basis technique is a viable approach to IBP reduction, of which potentially also synergies with existing implementations can be identified in the future. However, new conceptual ideas are needed to deal with the enormous intermediate expression swell with increasing complexity of reduction problems.

\section*{Acknowledgments}

This research was supported by the Deutsche Forschungsgemeinschaft (DFG,
German Research Foundation) under grant 396021762 -- TRR 257 ``Particle Physics Phenomenology after the Higgs Discovery.''



\providecommand{\href}[2]{#2}\begingroup\raggedright\endgroup


\end{document}